\documentclass[conference]{IEEEtran}
\usepackage{amsmath}
\usepackage{cite}
\usepackage{graphicx}
\usepackage{amssymb}
\usepackage{bm}
\usepackage{stfloats}
\usepackage{cases}
\usepackage{color}
\usepackage{algorithm}
\usepackage{float}
\makeatletter
\newenvironment{breakablealgorithm}
{
	\begin{center}
		\refstepcounter{algorithm}
		\hrule height.8pt depth0pt \kern2pt
		\renewcommand{\caption}[2][\relax]{
			{\raggedright\textbf{\ALG@name~\thealgorithm} ##2\par}%
			\ifx\relax##1\relax 
			\addcontentsline{loa}{algorithm}{\protect\numberline{\thealgorithm}##2}%
			\else 
			\addcontentsline{loa}{algorithm}{\protect\numberline{\thealgorithm}##1}%
			\fi
			\kern2pt\hrule\kern2pt
		}
	}{
		\kern2pt\hrule\relax
	\end{center}
}
\makeatother

\usepackage{algpseudocode}
\DeclareMathOperator*{\st}{s.t.}

\newtheorem{Remark}{\underline{Remark}}

\ifCLASSINFOpdf
\else
\fi
\begin{document}
%
\title{Optimal Online Transmission Policy for Energy-Constrained Wireless-Powered Communication Networks}
%
%
%
\author{\IEEEauthorblockN{Xian Li\IEEEauthorrefmark{2}, Xiangyun Zhou\IEEEauthorrefmark{3}, Derrick Wing Kwan Ng\IEEEauthorrefmark{4}, and Changyin Sun\IEEEauthorrefmark{2}}\\
	
	\IEEEauthorblockA{\IEEEauthorrefmark{2}School of Automation, Southeast University, Nanjing, China\\}
	\IEEEauthorblockA{\IEEEauthorrefmark{3} Research School of Engineering, The Australian National University, Canberra, ACT, Australia\\}
	\IEEEauthorblockA{\IEEEauthorrefmark{4}School of Electrical Engineering and Telecommunications, The University of New South Wales, Sydney, NSW, Australia\\}
	\IEEEauthorblockA{Email: seulixian@gmail.com, xiangyun.zhou@anu.edu.au, w.k.ng@unsw.edu.au, cysun@seu.edu.cn}}
\maketitle

\begin{abstract}
This work considers the design of online transmission policy in a wireless-powered communication system with a given energy budget. The system design objective is to maximize the long-term throughput of the system exploiting the energy storage capability at the wireless-powered node. We formulate the design problem as a constrained Markov decision process (CMDP) problem and obtain the optimal policy of transmit power and time allocation in each fading block via the Lagrangian approach. To investigate the system performance in different scenarios, numerical simulations are conducted with various system parameters. Our simulation results show that the optimal policy significantly outperforms a myopic policy which only maximizes the throughput in the current fading block. Moreover, the optimal allocation of transmit power and time is shown to be insensitive to the change of modulation and coding schemes, which facilitates its practical implementation.
\end{abstract}


%
\IEEEpeerreviewmaketitle
\section{Introduction}
Wireless-powered communication networks (WPCNs), which usually consist of a hybrid access point (H-AP) and several user equipments (UEs) \cite{Wu2017Survey}, have drawn significant attention recently. The system performance in terms of different metrics (e.g., throughput \cite{Ju2014ThroughputMaximization}, outage \cite{Chen2015HTC}, energy efficiency \cite{Wu2016TWC:EE}) for various scenarios (e.g., point-to-point \cite{Kim2016}, two-hop relaying \cite{Luo2016}, multiple-input and multiple-output (MIMO) \cite{Diamantoulakis2016}) have been thoroughly investigated. However, most existing works devoted their efforts to studying the system performance of only one time block (slot), where all the harvested energy is exhausted immediately without exploiting long-term energy storage. In practice, due to the variability of the communication channel quality, it is more reasonable to store part of or even all the harvested energy in the battery when the channel undergoes deep fading. Thus it is of great importance to study the transmission policy for optimizing long-term system performance with long-term energy storage capability. \

Some research efforts have been devoted to improving the long-term system performance. Considering two simple online transmission policies for a single-user WPCN, the limiting distribution of the stored energy at the UE as well as the outage performance of the system was investigated in \cite{Morsi2018}. In \cite{Zhou2016}, the data rate maximization problem of an orthogonal frequency division multiplexing (OFDM)-based WPCN was studied. To jointly optimize the subchannel allocation and the power allocation over time, an offline algorithm and an online algorithm were designed for the case of non-causal channel state information (CSI) and causal CSI, respectively. Considering the variation of the CSI and the evolution of the battery state over slots, the long-term system performance of a two-user WPCN in an infinite horizon was studied in \cite{Biason2017}. Based on the theory of Markov decision process, the optimal online policy was obtained to maximize the long-term system throughput. After that, the authors in \cite{Abd-Elmagid2016} extended this work to a full-duplex scenario where the H-AP transfers energy and receives information data simultaneously. The corresponding optimal online policy for the full-duplex case was obtained and the long-term performance gap between the full-duplex WPCN and the half-duplex WPCN was also discussed. However, the temporal correlation of the time-varying channels, which can be exploited to improve the system performance, was not considered in these works. Also, the H-AP in these works, e.g., \cite{Morsi2018,Zhou2016,Biason2017,Abd-Elmagid2016}, was assumed to equip with an infinite power supply and hence energy consumption of the system has not been a consideration in the previous studies.\

In this paper, we focus on the long-term throughput performance of a WPCN with limited system energy budget. More specifically, considering the H-AP with a finite amount of energy, we design an optimal online transmission policy to maximize the throughput over an infinite horizon. The contribution of the work lies in both the modeling and solution development of the throughput maximization problem. First, during problem formulation, the finite state Markov channel (FSMC) model is adopted to capture the temporal-correlation behavior of the fading channel. Moreover, practical aspects including circuit power consumption and efficiency of the power amplifier are considered to evaluate the total system energy consumption. Then, we formulate the problem as a constrained Markov decision process (CMDP) problem and solve it optimally via the Lagrangian approach, where a bisection search is introduced to update the corresponding Lagrange multiplier. Subsequently, the long-term system performance under various scenario is studied via numerical simulations. In particular, the impact of the system parameters on the system performance is thoroughly discussed, which provides practical insights on the design and implementation of the WPCN.\


\section{System model}\label{sec2}
\begin{figure}[h]
	\centering
	\includegraphics[scale=1]{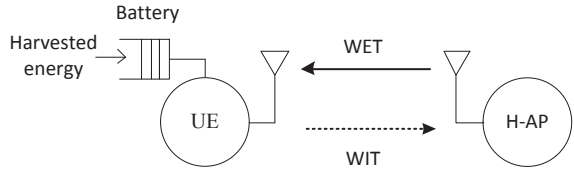}
	\caption{The system model of a WPCN.}
	\label{sys_mod}
\end{figure}

As shown in Fig. \ref{sys_mod},  we consider a WPCN consisting of a H-AP and a single-antenna UE in this paper. The H-AP is equipped with a directional antenna and the UE is driven by a rechargeable battery with maximum capacity $B_{\rm {max}}$. A time-correlated block fading channel is considered between the H-AP and the UE, where the channel power gain remains constant in a block but varies from one to another. In block $t \in{1,2,\cdots}$, the channel power gain is expressed as $H_{t}=\theta_td^{-\alpha}$, where $\theta_t$ is a random variable capturing the multipath fading, $d$ is the distance between the H-AP and the UE, and $\alpha$ is the path loss exponent. In each block, there is a wireless energy transfer (WET) period and a wireless information transfer (WIT) period. The UE first harvests energy from the H-AP and stores it in the battery during WET, and then transmits its data to the H-AP utilizing the energy stored in the battery during the following WIT. \

In this paper, we aim at maximizing the system throughput over an infinite horizon under a given energy budget constraint. This considered problem can be formulated in the framework of a CMDP which consists five elements: the system state space $\mathcal{S}$, the action space $\mathcal{A}$, the probability transition matrix $\mathcal {P}$, the reward function $r(\cdot )$, and the cost function $e(\cdot )$. In the following, detail descriptions of these five elements are provided.\

\subsection{System States}
For the considered system, the optimal policy is constructed at the H-AP based on the channel information and the battery information. We assume that in the current block, perfect CSI as well as the UE's battery information is available at the H-AP (In practice, this information can be acquired in the training phase at the beginning of each block). Correspondingly, in block $t$, the system state $\bm s_t\in \mathcal{S}$ consists the channel state $h_t\in \mathcal{H}$ and the battery state $b_t\in \mathcal{B}$, i.e., $\bm s_t=[h_t, b_t]$. Similar to the works in \cite{Mao2014,Li2018,Zhang2014,Sadeghi2008a}, quantized system state is considered in this paper. Specifically, the system state space $\mathcal{S}$ is expressed as $\mathcal{S}=\mathcal{H}\times \mathcal{B}$, where $\mathcal{H}\triangleq \{1,2,...,K\}$ and $\mathcal{B}\triangleq \{0,...,l,...,L\}$ define the set of channel state and battery state, respectively. The battery is at state 0 when the stored energy is exhausted.\

In practice, the channel in a communication system is generally time-correlated. As stated in \cite{Sadeghi2008a,Wang1995,Babich2000}, the time-varying behavior of the fading channel can be well captured by the FSMC model. Accordingly, in this paper, we separate the channel gain by a set of boundaries, i.e., $\bm \Gamma =\{{\Theta}_1,{\Theta}_2,...,{\Theta}_k,...{\Theta}_{K+1}\}\times d^{-\alpha}$, where ${\Theta}_k$ varies in an increasing order with ${\Theta}_1=0$ and ${\Theta}_{K+1}={\infty}$. In the $t$-th block, the channel state $h_t \in \mathcal{H}$ is said to be at state $k$ (i.e., $h_t=k$) if ${\Theta}_k{\leq}\theta_t<{\Theta}_{k+1}$. \

We assume that there is only an one-step channel state transition from block to block. Denoting $\pi _k$ as the steady state probability of the channel being at state $k$. With equiprobable
partition of the channel gain (this is a reasonable and commonly adopted technique in a FSMC model, cf.\cite{Sadeghi2008a,Li2018,Zhang2014}), i.e., $\pi _k=\frac{1}{K}, \forall k \in \{1,2,...,K\}$, the fading boundaries $\Theta_k$ can be obtained by solving the following equations:
\begin{equation}
\begin{split}
\pi _k&=\int_{{\Theta}_k}^{{\Theta}_{k+1}}\rho (\theta_t)d\theta_t=\frac{1}{K},\forall k\in\{1,2,...,K\},
\end{split}
\end{equation}
where $\rho (\theta_t)$ is the probability density function of the variable $\theta_t$. When channel is at state $k$, i.e., $h_t=k$, the quantized value of the channel gain is\
\begin{equation}\label{Eq_Exp_H}
\begin{split}
\bar{H_t}=\frac{\int_{\Theta_k}^{\Theta_{k+1}}H_t\rho(\theta_t){\mathrm{d}\theta_t }}{\int_{\Theta_k}^{\Theta_{k+1}}\rho(\theta_t){\rm{d}\theta_t }}=\frac{\int_{\Theta_k}^{\Theta_{k+1}}\theta_t d^{-\alpha}\rho(\theta_t){\mathrm{d}\theta_t }}{\pi_k}.
\end{split}
\end{equation}

Similarly, the available energy in the battery of the UE is discretized into $L$ quantum. Denote $Q$ as one energy quantum level of the battery, then the maximum capacity of the battery is $B_{\rm{max}}=LQ$. In the $t$-th block, the battery state is said to be at state $l$ (i.e., $b_t=l$) if $\lfloor \frac{B_t}{Q} \rfloor =l$, where $B_t$ is the available battery energy at the beginning of block $t$.\

\subsection{Actions, Reward, and Cost Functions}
At the beginning of each block, the H-AP makes a decision according to the current system state and reports it to the UE such that the system is well scheduled during the following WET and WIT procedure. The time duration of each block $T$ is divided into two orthogonal time slots: ${\tau}_t^{\rm E}$ for WET and ${\tau}_t^{\rm I}$ for WIT with ${\tau}_t^{\rm E}+{\tau}_t^{\rm I}\leq T$. Let $P_t^{\rm E}$ and $P_t^{\rm I}$ be the transmit power of the H-AP for WET and the transmit power of the UE for WIT, respectively. Then, the action adopted in block $t$ (denoted by $\bm {a}_t$) contains four elements, i.e., $\bm {a}_t=\{{\tau}_t^{\rm E},{\tau}_t^{\rm I},P_t^{\rm E},P_t^{\rm I}\}$. \

For a given system state, different actions come with different rewards and costs. In our work, we consider the throughput per block (defined as the data bits transmitted in one block) as the immediate reward and the energy consumption per block as the immediate cost. Denote the feasible action set at state $\bm s_t$ as $\mathcal{A}(\bm s_t)$. For a given state $\bm s_t$ and an action $\bm a_t \in \mathcal{A}(\bm s_t)$, the immediate reward, i.e., $r(\bm s_t, \bm a_t)~:~\mathcal{S}\times \mathcal{A}\rightarrow \mathbb{R}$, is defined as
\begin{equation}
\begin{split}
r(\bm s_t, \bm a_t)=\frac{\int_{\Theta_k}^{\Theta_{k+1}}{\tau}_t^{\rm I}W\text{log}_2\left (1+\frac{P_t^{\rm I}{\theta_td^{-\alpha}}}{\zeta \sigma ^2}\right )\rho(\theta_t){\rm{d}\theta_t }}{\pi_k}, \label{rewardeq}
\end{split}
\end{equation}
where $W$ is the bandwidth of the considered system, $\sigma ^2=N_0W$ is the thermal noise power (where $N_0$ is the noise power density), and the factor $\zeta$ characterizes the discrepancy between the achievable rate and the channel capacity due to the use of practical modulation and coding schemes \cite{Wu2016TWC:EE}. \

The corresponding immediate cost, i.e., $e(\bm s_t, \bm a_t)~:~\mathcal{S}\times \mathcal{A}\rightarrow \mathbb{R}$, is expressed as
\begin{equation}
e(\bm s_t, \bm a_t)=\frac{P_t^{\rm E}{\tau}_t^{\rm E}}{{\vartheta}_{\rm{AP}}}+P_{\rm {C_{AP}}}{\tau}_t^{\rm E}+e_t^{\rm IT}-e_t^{\rm {AC}},\label{consteq}
\end{equation}
where the first two terms capture the energy consumption at the H-AP and the last two terms describe the battery consumption at the UE. Specifically, $0<{\vartheta}_{\rm{AP}}<1$ is the power amplifier efficiency of H-AP. Hence the first term in \eqref{consteq} presents the energy consumption of the power amplifier during WET. $P_{\rm{C_{AP}}}$ is the circuit power at the H-AP. Hence the second term in \eqref{consteq} accounts for the energy consumption of the circuit during WET. For the battery consumption at the UE, 
\begin{equation}
e_t^{\rm IT}=\frac{P_t^{\rm I}{\tau}_t^{\rm I}}{{\vartheta}_{\rm U}}+P_{\rm {C_U}}{\tau}_t^{\rm I}
\end{equation}
stands for the energy consumption of the UE during WIT, where ${\vartheta}_{\rm U}$ and $P_{\rm {C_{U}}}$ denote the power amplifier efficiency and the circuit power at the UE, respectively. Finally,
\begin{equation}
e_t^{\rm {AC}}=\min \left({B_t+{\eta}G_{\rm A}P_t^{\rm E}{\tau}_t^{\rm E}\bar{H}_t,B_{\rm{max}}}\right)-B_t
\end{equation}
is the energy accumulated in the battery in block $t$, where $\eta$ is the energy conversion efficiency and $G_{\rm A}$ is the antenna gain at the H-AP during WET. Obviously, the value of $e_t^{\rm IT}-e_t^{\rm {AC}}$ can be either positive (battery consumption) or negative (battery accumulation). \

By the conservation of energy, both $r(\bm s_t, \bm a_t)$ and $e(\bm s_t, \bm a_t)$ are nonnegative. Since the available energy of the UE in block $t$ is limited by the current stored energy in the battery, the feasible action set at system state $\bm s_t$ can be given as: 
\begin{equation}\label{feasible_A}
\begin{split}
\mathcal{A}(\bm s_t)=&\{\bm a_t|{\tau}_t^{\rm E}+{\tau}_t^{\rm I}{\leq}T, {\tau}_t^{\rm E}{\geq}0, {\tau}_t^{\rm I}{\geq}0, {P_t^{\rm I}}\geq 0,\\
&0{\leq}{P_t^{\rm E}}{\leq}{P_{\rm {max}}^{\rm E}}, e_t^{\rm IT}{\leq}e_t^{\rm {AC}}+B_t\},
\end{split}
\end{equation}
where $P_{\rm {max}}^{\rm E}$ is the maximum transmit power of the H-AP.

\subsection{Transition Probabilities}
Denote the system state in block $t$ and $t+1$ as $\bm s_t$ and $\bm s_{t+1}$, respectively. For an adopted action $\bm a_t$, the transition probability from state $\bm s_t$ to state $\bm s_{t+1}$ can be given as
\begin{equation}
\begin{split}
\mathcal{P}(\bm s_{t+1} |\bm s_t,\bm a_t)&\overset{(\text{a})}=\mathcal{P}(h_{t+1} ,b_{t+1} |h_t,b_t,\bm a_t)\\
&\overset{(\text{b})}=\mathcal{P}(h_{t+1} |h_t)\mathcal{P}(b_{t+1} |h_t,b_t,\bm a_t),
\end{split}
\end{equation}
where (a) holds by definition and (b) holds for the independence of the channel state evolution from the battery state and the action. In the following, we calculate the channel state transition probability $\mathcal{P}(h_{t+1} |h_t)$ and the battery state transition probability $\mathcal{P}(b_{t+1} |h_t,b_t,\bm a_t)$, respectively.\

The channel state transition probability, which is closely related to the time-varying behavior of the channel gain, can be described by the level crossing rate $\Lambda(\Theta)$ \cite{Sadeghi2008a,Wang1995,Babich2000}, i.e., the average number of times that the instantaneous value of $\theta_t$ crosses a given level $\Theta$. Specifically, the channel state transition probability from state $h_t$ to $h_{t+1}$ can be approximated by the ratio of $\Lambda(\Theta)$ divided by the average number of blocks the value of $\theta_t$ falls in the interval associated with the state $h_t$. Similar to \cite{Li2018,Zhang2014,Sadeghi2008a,Wang1995,Babich2000}, we assume that the channel state transits between its adjacent state only (the validity of this  commonly-used assumption has been verified in \cite{Wang1995}). Then, the channel transition probabilities can be approximated as
\begin{align}
\mathcal{P}(h_{t+1}&=k+1 |h_t=k)\approx \frac{\Lambda(\Theta_{k+1})T}{\pi_k},\\
\mathcal{P}(h_{t+1}&=k-1 |h_t=k)\approx \frac{\Lambda(\Theta_{k-1})T}{\pi_k}, \\
\mathcal{P}(h_{t+1}&=k |h_t=k)\approx 1-\frac{\Lambda(\Theta_{k+1})T}{\pi_k}-\frac{\Lambda(\Theta_{k-1})T}{\pi_k}.
\end{align}

On the other hand, the battery state transition can be described as follows. If $b_{t+1} < L$, \begin{equation}
\mathcal{P}(b_{t+1} |h_t,b_t,\bm a_t)=\delta\{b_t + \lfloor \frac{e_t^{\rm AC}-e_t^{\rm IT}}{Q}\rfloor=b_{t+1}\},
\end{equation}
otherwise,
\begin{equation}
\mathcal{P}(L |h_t,b_t,\bm a_t)=\delta\{b_t + \lfloor \frac{e_t^{\rm {AC}}-e_t^{\rm IT}}{Q}\rfloor\geq L\},
\end{equation}
where $\delta{(\cdot)}$ is the indicator function.

\section{CMDP Formulation and The Optimal Policy}\label{sec_3}
In this section, we formulate the CMDP problem and provide the corresponding optimal solution.\

\subsection{Problem Formulation}
For a system in the long run, a policy $\bm \mu$ is a sequence of decision rules, i.e., $\bm \mu=\{\mu_1, \mu_2,...\}$, each in which is a function mapping from the system state $\bm s$ to the action to be taken, i.e., $\mu_t: \mathcal{S} \rightarrow \mathcal{A}$, $\forall t$. A policy $\bm \mu$ is said to be stationary if the decision rule in it is independent with time, i.e., $\mu_1=\mu_2=\cdots$. If a policy is stationary and deterministic, then it is called a pure policy. To model the imperfect operation of the system in Fig. \ref{sys_mod}, we introduce a factor $\lambda \in [0,1)$ to capture the probability that the system hardware survives from a operation failure in a block. Correspondingly, as described in \cite{Altman1998CRC:Constrained}, for an available stationary policy $\bm \mu$, the long-term throughput of the system can be defined as 
\begin{equation}\label{ltobj}
R(\bm s_0,\bm \mu)=(1-\lambda)\sum_{t=1}^{\infty} {\lambda}^t\mathbb{E}_{\bm {s_0}}^{\bm \mu}\{{r(\bm {s}_t,\bm {a}_t)}\},
\end{equation}
and the long-term energy cost of the system can be defined as
\begin{equation}\label{Ecfunc}
E(\bm s_0,\bm \mu)=(1-\lambda)\sum_{t=1}^{\infty} {\lambda}^t\mathbb{E}_{\bm {s_0}}^{\bm \mu}\{{e(\bm {s}_t,\bm {a}_t)}\}.
\end{equation}
When $\lambda$ approaches 1, the discounted functions defined in \eqref{ltobj} and \eqref{Ecfunc} converge to their expected average values \cite{Altman1998CRC:Constrained}, respectively, which are defined in the form of $\lim_{N \rightarrow \infty}\frac{1}{N}\sum_{t=1}^{N}{\lambda}^t\mathbb{E}_{\bm{s}_0}^{\bm \mu}\{{X_t(\bm {s}_t,\bm {a}_t)}\}, X \in\{r,e\}$, where $N$ is the number of blocks. Thus \eqref{ltobj} and \eqref{Ecfunc} can be interpreted as the expected average throughput and the expected average energy cost per block, respectively. \

In this paper, we aim at finding an optimal policy $\bm \mu ^\ast$ such that the long-term throughput is maximized under a given energy budget $E_{\rm th}$. This policy can be obtained through solving the following CMDP problem:
\begin{subequations}\label{prob_formu}
\begin{align}
\underset{{\bm \mu}}{\max}~~~~&R(\bm s_0,\bm \mu)\label{ThroughputOpt}\\
\st ~~~~&E(\bm s_0,\bm \mu){\leq}E_{\rm th}.\label{Econstraintprob1}
\end{align}
\end{subequations}

\subsection{The Optimal Policy}
As shown in \cite{Altman1998CRC:Constrained}, the CMDP problem in the form of \eqref{prob_formu} can be efficiently solved via the Lagrangian approach, whereby the CMDP problem is transferred into an equivalent unconstrained MDP problem. Accordingly, by introducing a non-negative Lagrangian multiplier $\beta$ for problem \eqref{prob_formu}, a new reward function $\widetilde r(\bm s,\bm a;\beta):~\mathcal{S}\times \mathcal{A}\times \mathbb{R}^+ \rightarrow \mathbb{R}$, can be constructed for the equivalent unconstrained MDP problem, where
\begin{equation}
\widetilde r(\bm s,\bm a;\beta)=r(\bm s,\bm a)-\beta e(\bm s,\bm a),
\end{equation}
and the corresponding Bellman's optimality equation is:
\begin{equation}\label{OptEq}
\begin{split}
J_\beta (\bm s)=\underset{\bm a \in \mathcal{A}(\bm s)}{\max} &\left \{(1-\lambda)\widetilde r(\bm s,\bm a;\beta)\right.\\
&\left.+\lambda \sum_{\bm s^\prime \in \mathcal{S}}\mathcal{P}(\bm s^\prime |\bm s,\bm a)J_\beta (\bm s^\prime) \right \},
\end{split}
\end{equation}
which can be efficiently solved via the Value Iteration Algorithm (VIA) \cite{Puterman2005:MDP} for any fixed $\beta$. Correspondingly, the optimal policy with a given $\beta$, i.e., $\bm \mu_{\beta}=\{\mu_{\beta}(\bm s), \forall \bm s\in\mathcal{S}\}$, can be determined by:
\begin{equation}\label{OptPolicy}
\begin{split}
\mu_\beta(\bm s)=\underset{\bm a \in \mathcal{A}(\bm s)}{\arg}{\max} &\left \{(1-\lambda)\widetilde r(\bm s,\bm a;\beta)\right.\\
&\left.+\lambda \sum_{\bm s^\prime \in \mathcal{S}}\mathcal{P}(\bm s^\prime |\bm s,\bm a)J_\beta (\bm s^\prime) \right\}.
\end{split}
\end{equation}

As described in \cite{Altman1998CRC:Constrained}, the optimal policy of a CMDP problem with a single constraint is composed of two pure policies, i.e., $\bm \mu_{\beta^-}$ and $\bm \mu_{\beta^+}$, with $\beta ^-$ and $\beta ^+$ as their associated Lagrangian multipliers, respectively. The policy $\bm \mu_{\beta^-}$ yields the highest energy cost $E^-$ that satisfies the energy constraint, while the policy $\bm \mu_{\beta^+}$ yields the lowest energy cost $E^+$ that breaks the energy constraint. Since $J_\beta (\bm s)$ is a monotonically non-increasing function of $\beta$ \cite{Beutler1985}, the value of $\beta^-$ and $\beta^+$ can be efficiently obtained via the bisection search method. With a randomized mixture of $\bm \mu_{\beta^-}$ and $\bm \mu_{\beta^+}$, the optimal policy of a CMDP problem can be given by: 
\begin{numcases}{\bm \mu^\ast=}
\bm \mu_{\beta^-}, &w.p. $q$\\
\bm \mu_{\beta^+}, &w.p. $1-q$
,\end{numcases}
where the mixing weight parameter $0\leq q\leq 1$ can be obtained via solving equation $E_{\rm th}=qE^-+(1-q)E^+$. \

Correspondingly, the procedures for solving problem \eqref{prob_formu} is described in Algorithm \ref{alg1}. Since the optimal policy consists of two pure policies, both of
which are irrelevant to time sequence. In Algorithm \ref{alg1}, we drop the subscript ``$t$'' for convenience. Specifically, initializations are performed in line 1, where $n$ and $\varepsilon_{\beta}$ are the iteration sequence and the error bound for updating $\beta$, respectively. The initial value of $\beta^+$ is specified in an incremental method, i.e., increasing the initial value of $\beta^+$ until that the corresponding long-term system energy cost exceeds $E_{\rm th}$. The VIA is conducted to solve the equivalent unconstrained MDP problem in line 4 and the Lagrangian multiplier $\beta$ is updated via bisection search in lines 5-13. Finally, with the obtained policy $\mu_{\beta^-}(\bm{s})$ and  $\mu_{\beta^+}(\bm{s})$, the mixing weight $q$ and the optimal policy are obtained in line 17 and line 18, respectively.\

For the implementation of VIA, the candidate actions at each state are quantized. Specifically, $\tau^{\rm E}$, ${\tau}^{\rm I}$, $P^{\rm E}$, and $P^{\rm I}$ are discretized into levels of $V_{\tau}^{\rm E}$, $V_{\tau}^{\rm I}$, $V_{P}^{\rm E}$, and $V_{P}^{\rm I}$, respectively. Since the update of $\beta$ is independent from the action space and the channel state space, the computational complexity of Algorithm \ref{alg1} is $\mathcal{O}(\frac{1}{1-\lambda}\log(\frac{1}{1-\lambda})V_{\tau}^{\rm E}V_{\tau}^{\rm I}V_{P}^{\rm E}V_{P}^{\rm I}|\mathcal{S}|^3)$ \cite{Littman1995}.\

\begin{Remark}
	In this paper, we obtain the optimal online policy for the CMDP problem \eqref{prob_formu} for the case of single UE. For the case of $M>1$ UEs, the corresponding tuple of the CMDP can be constructed as follow (here, we use the subscript ``$m$'' to denote the elements of the $m$-th UE): the system space $\bar{\mathcal{S}}$ can be expressed as $\bar{\mathcal{S}}=\mathcal{S}_1 \times \mathcal{S}_2 ...\times \mathcal{S}_m ...\times \mathcal{S}_M$, where $\mathcal{S}_m=\mathcal{H}_m \times \mathcal{B}_m$ is the system state space of the $m$-th UE and ``$\times$'' is the Cartesian product; the action space $\bar{\mathcal{A}}$ can be expressed as $\bar{\mathcal{A}}=\mathcal{A}_1 \times ... \mathcal{A}_m ...\times \mathcal{A}_M$, where $\mathcal{A}_m$ presents the action space of the $m$-th UE and is in the form of \eqref{feasible_A}; for an action $\bar{\bm a}_t=[\bm a_{1,t},...,\bm a_{m,t},...,\bm a_{M,t}]$ adopted at state $\bar{\bm s}_t=[\bm s_{1,t},...,\bm s_{m,t},...,\bm s_{M,t}]$, the immediate reward and the immediate cost of the system can be defined as $\bar{r}(\bar{\bm s}_t,\bar{\bm a}_t)=\sum_{m=1}^{M}r(\bm s_{m,t}, \bm a_{m,t})$ and $\bar{e}(\bar{\bm s}_t,\bar{\bm a}_t)=\sum_{m=1}^{M}e(\bm s_{m,t}, \bm a_{m,t})$, respectively; the system state transition probability matrix can be expressed as $\mathbb{P}=\mathbb{P}_1 \otimes... \mathbb{P}_m ...\otimes\mathbb{P}_M$, where $\mathbb{P}_m=\left[\mathcal{P}(\bm s_{m,t+1} |\bm s_{m,t},\bm a_{m,t})\right]$ is the system state transition probability matrix of the $m$-th UE and $\otimes$ is the Kronecker product. Based on this tuple, the CMDP problem for the multi-user case can be constructed and the corresponding optimal online policy can be obtained similarly via Algorithm 1.
\end{Remark}

\begin{breakablealgorithm}	\caption{The Optimal Policy for the CMDP \eqref{prob_formu}}
	\label{alg1}
	\begin{algorithmic}[1] 
		\State Set $n = 0$, $\beta^-$ = 0, $\beta^+$, $\beta^0 =\beta^-$, specify $\varepsilon_{\beta}>0$.
		\Repeat
		\State Set $\beta = \beta^n$ and $n=n+1$.
		\State For a given $\beta$, obtain the optimal policy $\bm \mu_\beta=\{\mu_{\beta}(\bm{s}),\forall \bm{s}\in\mathcal{S}\}$ via VIA.
		\State Compute the stationary distribution $\Psi(\bm s)$ induced by $\bm \mu_\beta=\{\mu_{\beta}(\bm{s}),\forall \bm{s}\in\mathcal{S}\}$.
		\If {$\sum_{\bm s \in \mathcal{S}}\Psi(\bm s)e(\bm s, \mu_\beta(\bm s)) > E_{\text{th}}$}
		\State $\beta^{n+1}=\frac{\beta^+ + \beta^n}{2}$.
		\State $\beta^- = \beta^n$.
		\Else
		\State $\beta^{n+1}=\frac{\beta^- + \beta^n}{2}$.
		\State $\beta^+ = \beta^n$.
		\EndIf
		\Until $|\beta^{n+1}-\beta^n|<\varepsilon_{\beta}$.	
		\State Find the policies $\bm \mu_{\beta^-}=\{\mu_{\beta^-}(\bm{s}),\forall \bm{s}\in\mathcal{S}\}$ and $\bm \mu_{\beta^+}=\{\mu_{\beta^+}(\bm{s}),\forall \bm{s}\in\mathcal{S}\}$ with obtained $\beta^-$ and $\beta^+$, respectively.
		\State Compute the stationary distribution $\Psi_{\beta^-}(\bm s)$ and $\Psi_{\beta^+}(\bm s)$ induced by $\bm \mu_{\beta^-}$ and $\bm \mu_{\beta^+}$, respectively.
		\State Compute
		\begin{equation}
		R_{\beta^-}=\sum_{\bm{s} \in \mathcal{S}}\Psi_{\beta^-}(\bm{s})r(\bm{s},\mu_{\beta^-}(\bm{s})),
		\end{equation}
		\begin{equation}
		R_{\beta^+}=\sum_{\bm{s} \in \mathcal{S}}\Psi_{\beta^+}(\bm{s})r(\bm{s},\mu_{\beta^+}(\bm{s})),
		\end{equation}
		\begin{equation}
		E_{\beta^-}=\sum_{\bm{s} \in \mathcal{S}}\Psi_{\beta^-}(\bm{s})e(\bm{s},\mu_{\beta^-}(\bm{s})),
		\end{equation}
		\begin{equation}
		E_{\beta^-}=\sum_{\bm{s} \in \mathcal{S}}\Psi_{\beta^+}(\bm{s})e(\bm{s},\mu_{\beta^+}(\bm{s})).
		\end{equation}
		\State Compute $q$ by solving $E_{\text{th}}=qE_{\beta^-}+(1-q)E_{\beta^+}$.
		\State Obtian the optimal reward $R=qR_{\beta^-}+(1-q)R_{\beta^+}$ and the optimal policy
		\begin{numcases}{\bm \mu^\ast=}
		\bm \mu_{\beta^-}, &w.p. $q$ \\
		\bm \mu_{\beta^+}, &w.p. $1-q$ 
		\end{numcases}	
	\end{algorithmic}
\end{breakablealgorithm}

\section{Simulation Results}\label{sec_4}
In this section, numerical simulations are provided for evaluating the long-term throughput performance of the system. For the practicality of RF energy transfer, a Rician fading channel is considered between the H-AP and the UE \cite{Zeng2015a,Zhao2018}. Correspondingly, the PDF of $\theta_t$ is given by
\begin{equation}
\rho (\theta_t)=\frac{1}{2\varrho ^2}e^{\frac{-(\theta_t +\varsigma^2)}{2\varrho ^2}}I_0\left (\frac{\sqrt \theta_t \varsigma}{\varrho ^2}\right ),
\end{equation}
where $I_0$ is the modified Bessel function of the zero-th order, $2\varrho ^2$ and $\varsigma^2$ are the parameters representing the power of multi-path and line-of-sight, respectively. Moreover, the level crossing rate $\Lambda(\Theta_\text{b})$ is\cite{Babich2000}
\begin{equation}
\begin{split}
\Lambda(\Theta)=\sqrt{\frac{2\pi(1+\kappa)\Theta}{\bar{\theta}}}f_De^{-(\kappa+\frac{1+\kappa}{\bar{\theta}}\Theta)}I_0(2\sqrt\frac{\kappa(1+\kappa)\Theta}{\bar{\theta}}),
\end{split}
\end{equation}
where $f_D$ is the maximum Doppler shift of the channel, $\bar{\theta}=2\varrho ^2+\xi^2$ is the local-mean fading power and $\kappa=\frac{\xi^2}{2\varrho ^2}$. Accordingly, practical channel parameters setting in \cite{Babich2000} is considered in simulations, where the number of channel states is selected as $K$ = 3, $f_D$ is set as 1.34 Hz, and the block duration is set as $T$ = 16 ms, respectively.\

Similar to \cite{Biason2017}, we focus on the case of small devices and express the battery size as a function of the reference value $B_{\rm {ref}}=10^{-3}\times T$ J. Unless otherwise stated, the maximum battery capacity is set as $B_{\rm max}=10B_{\rm ref}$. On the other hand, extensive simulations (not shown here) have revealed that the accuracy of results is guaranteed when $\varepsilon_{\beta}=10^{-4}$ and $Q=B_{\rm{ref}}$. Other important parameters used in simulations are listed in Table \ref{table1}. Moreover, to show the superiority of the optimal policy, the myopic policy which maximizes the throughput in only the current block is used as the benchmark. For legibility, in the simulation results, we mark the optimal policy and the myopic policy as ``Optimal'' and ``Myopic'', respectively.\

\begin{table}
	\centering
	\caption{Parameters Setting}
	\label{table1}
	\begin{tabular}{|c|c|c|c|c|c|}
		\hline
		$P_{\rm {max}}^{\rm E}$        & 10 W  &$\alpha$ &2.8     &$P_{\rm {C_{AP}}}$       &500 mW\\
		\hline
		$P_{\rm {C_{U}}}$        &5 mW   &${\vartheta}_{\rm {AP}}$ & 0.9   &${\vartheta}_{\rm U}$    &0.9 \\
		\hline
		$\eta$             & 0.95  &$\lambda$&0.9 &$G_A$             & 8 dBi \\
		\hline
		$\zeta$            & 1     &$W$                &2 kHz   &$N_0$       &-164 dBm/Hz\\
		\hline
		$\varsigma^2$      & 0.75 &$\varrho ^2$ & 0.125 & &\\
		\hline
	\end{tabular}
\end{table}

\begin{figure}[h]
	\centering
	\includegraphics[scale=0.56]{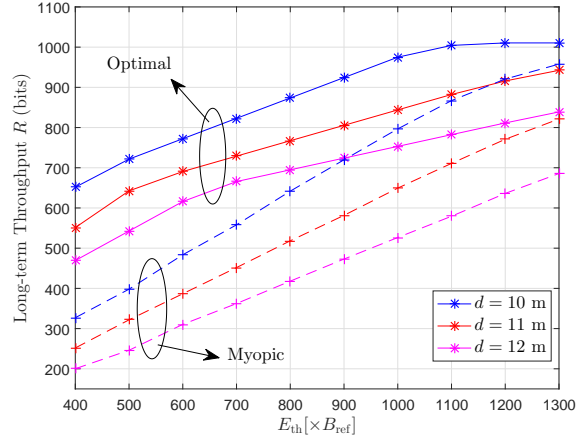}
	\caption{The long-term throughput versus the system energy budget $E_{\rm th}$.}
	\label{L2}
\end{figure}

\begin{figure}[!h]
	\centering
	\includegraphics[scale=0.56]{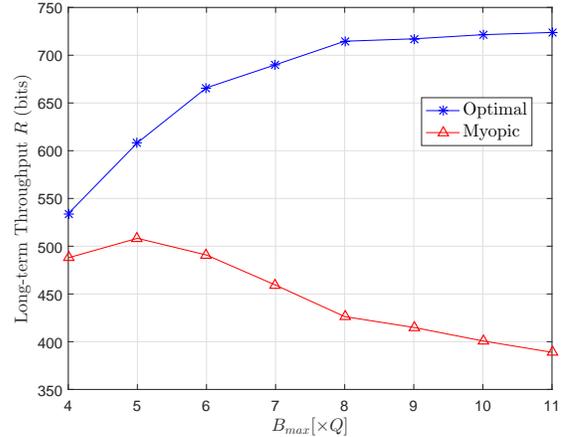}
	\caption{The long-term throughput versus the maximum battery
		capacity $B_{\rm max}$.}
	\label{L3}
\end{figure}

To investigate the impact of the energy budget and the communication distance on the system throughput performance, we first depict the long-term throughput as as a function of the energy budget $E_{\rm th}$ for different value of $d$. As shown in Fig. \ref{L2}, the optimal policy outperforms the myopic policy in all the considered cases. The long-term throughput is shown to be increased with $E_{\rm th}$. This is because that a larger $E_{\rm th}$ means more available energy budget. Due to the limitation of transmit power and the battery capacity, the system performance becomes saturated when $E_{\rm th}$ is exceedingly large (see the case of $d$=10 m). On the other hand, since the signal attenuations during WIT and WET are decreasing functions of the communication distance. As expected, the long-term throughput is shown to be reduced with $d$. \

The maximum battery capacity $B_{\rm max}$, which limits the maximum available energy at the UE in each block, is expected to have an impact on the system performance. Hence, in Fig. \ref{L3}, we investigate the long-term system throughput with varying $B_{\rm max}$. Here, we set $E_{\text{th}}=500B_{\text{ref}}$ and $d=10$ m. As shown in the figure, the long-term throughput with the optimal policy increases with $B_{\rm max}$. In fact, a larger $B_{\rm max}$ means a higher ability to handle the fluctuation of the channel state. As $B_{\rm max}$ grows, the performance gain becomes saturate due to the limitation of $E_{\rm th}$. However, the myopic policy shows a different trend. With the growth of $B_{\rm max}$, the corresponding long-term throughput first increases and then decreases. This is due to the fact that the myopic policy operates sequentially from block to block and exhausts the battery's energy as much as possible to maximize the current system throughput, which results in a trade-off on $B_{\rm max}$. Nevertheless, compared to the myopic policy, considerable improvement can be observed when the optimal policy is adopted.\

\begin{figure}[h]
	\centering
	\includegraphics[scale=0.56]{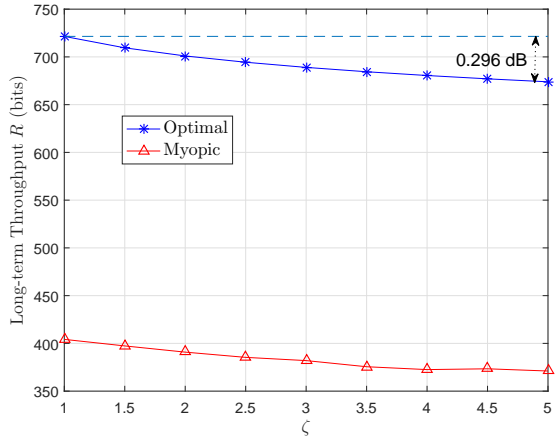}
	\caption{The long-term throughput versus the gap factor $\zeta$.}
	\label{L4_1}
\end{figure}

\begin{figure}[h]
	\centering
	\includegraphics[scale=0.56]{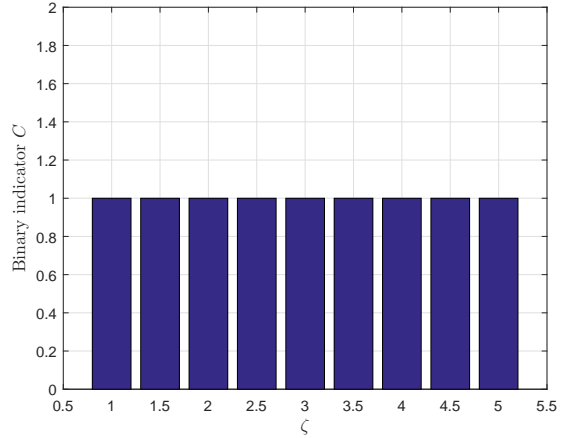}
	\caption{The binary indicator $C$ versus the gap factor $\zeta$.}
	\label{L4_2}
\end{figure}

\begin{figure}[h]
	\centering
	\includegraphics[scale=0.56]{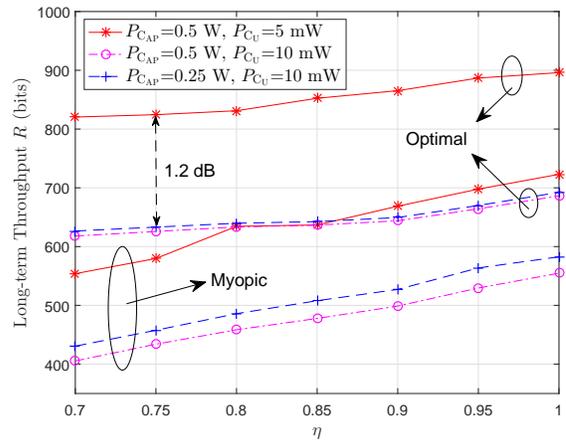}
	\caption{The long-term throughput versus the energy conversion efficiency $\eta$ with different circuit power $P_{\rm C_{AP}}$ and $P_{\rm C_{U}}$.}
	\label{L5}
\end{figure}

As stated in \eqref{rewardeq}, the factor $\zeta$ is used to capture the impact from the practical modulation and coding schemes. In Fig. \ref{L4_1}, we depict the long-term throughput as a function of $\zeta$ with $E_{\text{th}}=500B_{\text{ref}}$ and $d=10$ m. As shown in the figure, compared with the myopic policy, a high system performance gain is achieved when the optimal policy is adopted. Moreover, the long-term throughput is shown to be slightly decreased with the increasing $\zeta$. For example, with rising $\zeta$ from 1 to 5 (about 7 dB), the long-term throughput performance for the optimal policy drops only about 0.296 dB. On the other hand, the impact of $\zeta$ on the optimal policy is investigated in Fig. \ref{L4_2}. Here, we take the optimal policy with $\zeta=1$ (i.e., $\bm \mu_{\zeta=1}^\ast$) as the reference policy and use a binary indicator $C$ to identify the variation of the optimal policy with $\zeta$. Specifically, denote the optimal policy with $\zeta^\prime$ as $\bm \mu_{\zeta^\prime}^\ast$, then $C=1$ if $\bm \mu_{\zeta^\prime}^\ast$ is identical to $\bm \mu_{\zeta=1}^\ast$. Otherwise, $C=0$. As demonstrated in Fig. \ref{L4_2}, the value of $C$ equals to 1 and remains unchanged for different values of $\zeta$, which implies that the optimal policy is irrelevant to the practical implementation of the modulation and coding schemes.\

Lastly, the impact of the energy conversion efficiency and the circuit power on the system performance is investigated in Fig. \ref{L5}. Here we set $E_{\text{th}}=500B_{\text{ref}}$ and $d=8$m. As can be observed, the long-term throughput grows with the increasing of $\eta$. This is due to the fact that more available energy can be harvested at the UE with higher energy conversion efficiency. On the other hand, although $P_{\rm C_{AP}}$ dominates the circuit power of the whole system, the system performance is shown to be more sensitive to $P_{\rm C_{U}}$ rather than $P_{\rm C_{AP}}$. Specifically, with the optimal policy, the long-term throughput achieves a performance gain of 1.2 dB at $\eta=0.75$ when $P_{\rm C_{U}}$ decreases 3 dB (from 10 mW to 5 mW), but is almost unchanged when $P_{\rm C_{AP}}$ drops from 0.5 W to 0.25 W. In practice, this intrigues an prior effort on cutting down the circuit power consumption at the UE rather than at the H-AP.\

\section{Conclusion}\label{sec_5}
In this paper, we studied the problem of designing the optimal online policy in an energy-constrained WPCN to manage the transmit power and time durations for both WET and WIT over time-correlated fading channels. Aiming at maximizing the system long-term throughput with a limited energy budget, we formulate the transmission policy design as a CMDP problem, which was later transformed into an equivalent unconstrained MDP problem and solved via the Lagrangian approach. Numerical results showed that the long-term system performance is closely related to the total energy budget, the battery capacity, the communication distance, the energy conversion efficiency, and the circuit power of the system. 
For instance, the circuit power consumption at the UE has a stronger impact on the system performance than that at the H-AP. Also, the optimal policy was shown to be independent of the choices of modulation and coding schemes.

\section*{Acknowledgment}
This work was supported in part by the National Natural Science Foundation of China (U1713209, 61520106009, 61533008, 61573103) and in part by CSC. The work of X. Zhou was supported by the Australian Research Council's Discovery Projects Funding Scheme (Project number DP170100939). The work of D. W. K. Ng was supported by the Australian Research Council’s Discovery Early Career Researcher Award (DE170100137).

\ifCLASSOPTIONcaptionsoff
  \newpage
\fi

\bibliographystyle{IEEEtran}
\bibliography{IEEEabrv,MyRefLib}


\end{document}